# OPTIMIZED FILLING OF THE ALS STORAGE RING

Jan Pusina, Lawrence Berkeley National Laboratory, 80-101, Berkeley, CA, 94720

Abstract

The Storage Ring at the Advanced Light Source is capable of being filled in selectable buckets. Typically, 276 consecutive buckets out of 328 possible are selected. It has been found that the amount of charge injected is not uniform bucket-to-bucket, sometimes varying by 100% or more. This variability causes problems with beam stability, and it may shorten the beam lifetime. We have devised a feedback system that allows for beam filling in an orderly manner, yielding bunch heights uniform to within 17%. At fill time, multiple bunches are injected into the Storage ring first, then, using feedback, selected single bunches are used to top off the fill.

## 1 INTRODUCTION

The Storage Ring beam is monitored by a BPM in summing mode and wired into the control room to a Tektronix TDS 784A oscilloscope. A GPIB interface connects the scope signal to a Labview program, Ogetwave.vi[1], running on a PC. This program detects the bunch stream, processes the information, and periodically stores the file on a server. Another program written in Delphi, SRInject.exe, runs simultaneously, reading the files, and further analyzing and processing the bunch data. This program contains an algorithm which uses the individual bunch heights as feedback and selectively fills each of them according to how much charge is required to bring them up to an even height while filling the storage ring to a current of 393 mA. Then a "camshaft" bucket is automatically filled to 10mA in the dead space after the bunch train. By this time this last step is finished, it is assumed the multibunch buckets have decayed a few milliamps, bringing the total current to 400mA.

## 2 FILLING THE STORAGE RING

The ring is filled using a graphic method of bucket selection (see Fig. 3). The start and end buckets are specified, as well as the pattern of spacing and cycles that are made necessary by the 8 nanosecond spacing of gun bunches compared with the 2 nanosecond spacing of storage ring buckets. The 276 bucket filling sequence is begun with target buckets 1, 13, 25. . . and intervening numbers filled in by the 3 gun bunches. I.e., when bucket 1 is targeted, buckets 5 and 9 are automatically filled, because of the three gun bunches. After the first part of the cycle is complete and target bucket 265 has been reached, the pattern starts over with 2, 14, 26, . . ., then 3, 15, 27, . . ., and finally 4, 16, 28, . . .. Thus all buckets are filled once. In the case of non-optimized fills, this procedure would run freely, repeating as many times as necessary to achieve 393 mA, starting and ending at unpredictable target buckets. Since it would not have filled each bucket the same number of times, and the fact that each bucket would have received unpredictable amounts of beam on each shot, the fill would be uneven. See Figure 2.

## 3 OPTIMIZATION

The solution to the uneven fill is to regulate the amount of charge in each bucket. One problem immediately encountered when observing the oscilloscope trace and the digitized version of it on SRInject.exe, is that there is a distortion of the heights of isolated bunches, or bunches that are significantly higher than their neighbors, apparently due to a cabling mismatch. At this time, no hardware solution of this problem has been found, so a formula for compensation of this error has been derived in software. This formula makes use of the fact that the increased height of a given bunch is inversely proportional to the height of the bunch immediately preceding it. If the preceding bunch is zero, the height of the bunch is greater by an error, e, and the compensated height, $I_c$, of the bunch in question is calculated as

$$I_c = I_1 - e\left(1 - \frac{I_0}{I_1}\right),$$

where $I_0$ and $I_1$ are two consecutive bunch heights.

---

[1] Jones, Orland "OGETWAVE.EXE", ALS note, 2001.

At fill time, the Storage Ring current has decayed by about one halflife, i.e., to about 200 mA. At this point the standard multibunch fil l is executed for N cycles, i.e., each of the 276 buckets is targeted N times. Next, a determination is made whether there is room for another cycle. If not, the gun output is reduced so that the next cycle fills to just under the nominal fill point of 393 mA, or a threshold level ($V_t$) of 1.42 mA per bucket. To derive the proper gun output, its output was graphed as a function of grid bias. The graph is a typical triode curve (see Figure 1.) To obtain the reduced gun output, y, the gun grid bias is increased until

$$y \approx \frac{(V_t - V_{max})}{dI/dT},$$

where $V_t - V_{max}$ is the difference between the threshold voltage and maximum bunch voltage, and $dI/dT$ is the fill rate.

Once the fill has been extended such that any bucket is $>= V_t - 0.5 dIdT$, the single bunch optimization, or topoff begins. In this case, the heights of the buckets are recorded, and all buckets less than $V_t - 0.5 dIdT$ are targeted for a single bunch injection. Assuming the buckets to be of unequal charge, the last step is repeated until all buckets are within $V_t - 0.5 dIdT$. The gun output may be optionally lowered in the final iteration, as in multibunch fill, to make the topoff smoother. Fills with and without optimization and their spectra are shown in Figure 2.

The optimization procedure can be executed in a 3-button operation: Multibunch Prefill, Optimization with Single Bunch Topoff, and Single Bunch Camshaft fill. The simplification of the procedure with just a few strokes minimizes operator errors as the fill is repeated throughout the week.

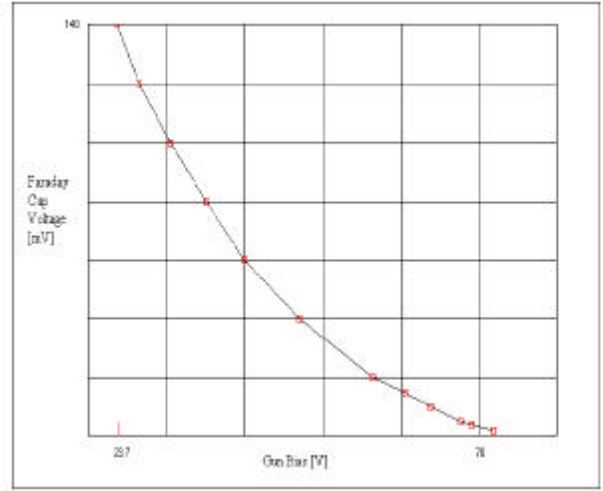

Figure 1: Gun output vs. Grid Bias

## 4 THE CAMSHAFT BUCKET

In the final phase of filling, the feedba ck scheme described above is also used to inject a camshaft bunch to a value of 10 mA. This bunch is used for precise timing of certain experiments, e.g., chemical dynamics, but its separation from and unequal charge with the other bunches is transparent t o other users.

All phases of filling can be done while observing a dynamic display of the buckets being filled as a bar graph, with the possibility of selecting an individual bucket and reading its height.

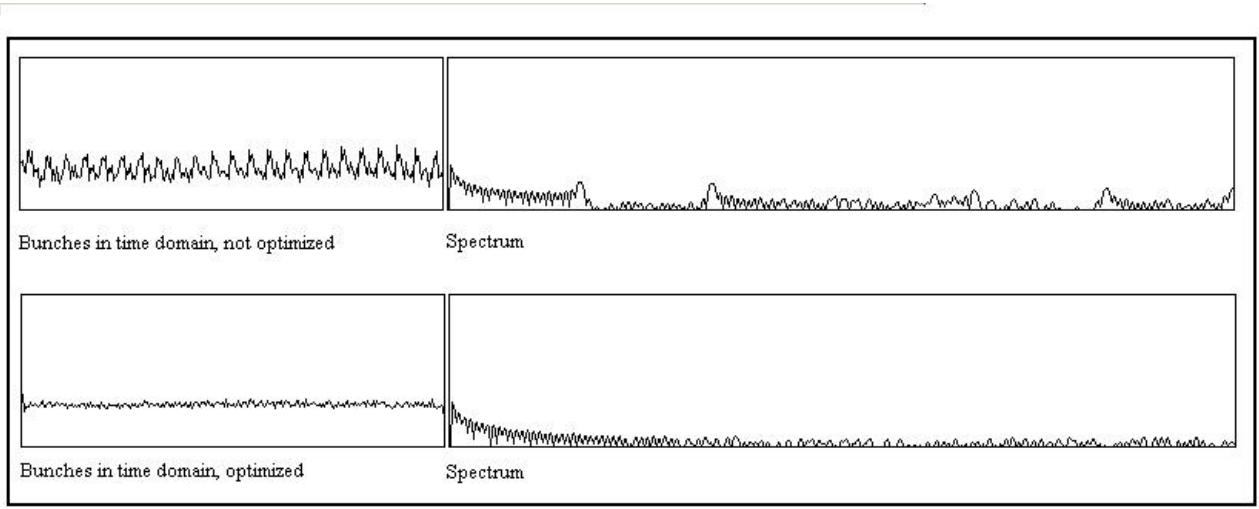

Figure 2: Bunches and Spectra

Figure 3: SRInject.exe